\documentclass[preprint,12pt]{elsarticle}
\usepackage{amssymb}
\usepackage{hyperref}
\journal{Physica A}

\usepackage{color}

\graphicspath{
{FIG_CAC_18/}
}
\begin{document}
\begin{frontmatter}
\title{Graph-based era segmentation of international financial integration}
%
% Single address.
% ---------------
%\name
\author{C\'ecile Bastidon$^{(1)}$, Antoine Parent$^{(2)}$, \\ Pablo Jensen$^{(3)}$, Patrice Abry$^{(3)}$, Pierre Borgnat$^{(3)}$ %\thanks{Work conducted in the CAC (Cliometrics \& Complexity group) at IXXI (Institut Rh�nalpin des Syst�mes Complexes), and supported by the GRAPHSIP project (ANR-14-CE27-0001-02) and the  ACADEMICS Grant given by the IDEXLYON project of the Universit\'{e} de Lyon, as part of the "Programme Investissements d'Avenir" ANR-16-IDEX-0005.}
}
\address{ 
$^{(1)}$ LEAD, Universit\'e de Toulon, France;  {\tt bastidon@univ-tln.fr}\\
$^{(2)}$ LAET, Sciences Po Lyon (UMR CNRS 5593), \& OFCE, France  ; {\tt antoine.parent@sciencespo-lyon.fr}\\
$^{(3)}$ Univ Lyon, ENS de Lyon, Univ Claude Bernard, CNRS, Laboratoire de Physique, F-69342 Lyon, France ;
{\tt \{pablo.jensen,patrice.abry,pierre.borgnat\}@ens-lyon.fr}
}

%
%\maketitle
%
\begin{abstract} % 150 mots max, ici 133
Assessing world-wide financial integration constitutes a recurrent challenge in macroeconometrics, often addressed by visual inspections searching for data patterns. Econophysics literature enables us to build complementary, data-driven measures of financial integration using graphs. The present contribution investigates the potential and interests of a novel 3-step approach that combines several state-of-the-art procedures to i) compute graph-based representations of the multivariate dependence structure of asset prices time series representing the financial states of 32 countries world-wide (1955-2015); ii) compute time series of 5 graph-based indices that characterize the time evolution of the topologies of the graph; iii) segment these time evolutions in piece-wise constant eras, using an optimization framework constructed on a multivariate multi-norm total variation penalized functional. 
The method shows first that it is possible to find endogenous stable eras of world-wide financial integration.
Then, our results suggest that the most relevant globalization eras would be based on the historical patterns of global capital flows, while the major regulatory events of the 1970s would only appear as a cause of sub-segmentation.    
\end{abstract}
\end{frontmatter}

\section{Introduction}
\label{sec:intro}

The scope of the present work is to combine data processing methods and graph-based representation of dependences in economic time series, in order to assess international financial integration in a historical perspective.
Financial integration means that the indices of equity markets of different countries should be highly correlated time series,
because, economically speaking, it would mean that the law of one price holds well from one country to another.
While the classical view in macroeconomics about international financial integration is based on the search for, and the interpretation of, a ``U-shaped curve" (e.g., \cite{o1999globalization,obstfeld2004global}), it remains a challenge to determine relevant eras of stable financial integration. 

Indeed, the facts and the dynamics of financial integration differs from one country to another.
For instance, there are two possible hypotheses as regards of the relevant causes and periodization of the history of international financial integration: a periodization based mainly on the {institutional framework} of international capital markets vs. an alternative periodization based on {private (or maket) decisions}, associated to actual changes in capital flows between countries, hence modifying the dependences between price series).
To make the matter more complex, delays between the possible causes and the change in financial integration may differ significantly from one country to another, and, obviously, regulatory shocks do not take place at the same date in different countries. 
%We show here that it is possible to escape these arbitrary decisions by looking globally at the structure of dependences between price indicators.

In this context, the present work aims to achieve an endogenous determination of periods, or eras, of international financial integration, delineated by critical changes in the dependence structure of global equity markets. 
Alternatively to the existing proposed chronologies of financial integration (following either hypotheses), this determination can be obtained directly from the indices of equity markets of different countries. For that, we estimate first what is the structure of dependences between these. Stable eras of financial integration should appear as periods were the dependences are not changing too much. 
Still, assessing local stationarity directly on the sliding estimates of the covariance structure would result into a too stringent procedure requesting that correlations between \emph{all pairs} of countries remain constant along time. 
Elaborating on ideas in \cite{mantegna1999hierarchical,bastidon2016form}, we propose to keep only a global description
of the financial integration by using a time-dependent graph representation of financial networks. Indeed, it was shown first in the seminal work \cite{mantegna1999hierarchical} that correlations between financial markets can be studied by keeping only the Minimum Spanning Tree on the correlation distance matrix, because it contains a meaningful economic taxonomy of the markets. 
It was used with great interest in more recent works, eg. \cite{TUMMINELLO201040,PhysRevE.84.026108,10.1371/journal.pone.0194067}.
This econophysics literature aims to study and manage portfolios, while here we consider equity markets at country level, in order to study international financial integration. From the graph-based representation of the evolving dependences between equity markets, a graph-based segmentation procedure is developed, which puts the focus on the global and topological structure of the graph, rather than on pairwise relations. This procedure will leverage the multivariate segmentation methods studied in~\cite{Frecon2014}, applied simultaneously to time-dependant features extracted from the graph representations of the dependences between equity markets.

% Thus, a challenge in the assessment of global financial integration in historical perspective is to determine relevant eras of stable financial integration.

\subsection{Related works} 

\subsubsection{Graph representation of dependences of financial networks}
The dependences between price indicators will be studied in the present work using a dependency graph. The starting point is the econophysics literature on securities portfolios, based on the groundbreaking contribution by Mantegna \cite{mantegna1999hierarchical,PhysRevE.62.R7615}, where he develops such an approach by computing a topological representation of financial networks, based on the common component of asset prices dynamics. The underlying principle is very close to the law of the one price of financial integration price indicators. 
From this representation, it is possible to study, e.g., the correlation structure of stock return time series \cite{TUMMINELLO201040}, the various time-scales in correlation and correlation-based graphs of  worldwide stock markets \cite{PhysRevE.84.026108}, or even cointegration or Granger causality \cite{10.1371/journal.pone.0194067}.
However, while the econophysics literature aims to quantify risk diversification in portfolio management over various periods of time, the objective in the present work is, as in \cite{bastidon2016form}, the assessment of international financial integration in a historical perspective. For that, a different point of view is adopted and most importantly a segmentation method that works globally over a long time span, will be needed.

\subsubsection{Data-driven based segmentation.} 
Alternative to the proposed approach, inspired  by econophysics and macroeconomics, data-driven signal processing approaches have been proposed to segment times series into \emph{stationary} periods. Also, segmentation of multivariate times-series according to their covariance structure is a topic that is recurrent in multivariate
signal processing, be it, e.g., for measurements in sensor network (e.g., weathers sensors), for analysis of financial data \cite{Xuan:2007:MCD:1273496.1273629}, for internet traffic data \cite{Gibberd_icassp2015_6854087}, or even for car sensors \cite{DBLP:journals/corr/HallacVBL17}.
The generic strategy consists in first the estimation of multivariate covariance structures from data (e.g., see classical methods in \cite{hastie2001elements}) followed by an a posteriori segmentation into periods where the covariance structure can be regarded as stable, or \emph{locally stationary}. 
Such covariance (or graphical models) segmentation can be achieved either using Bayesian frameworks (e.g., \cite{Xuan:2007:MCD:1273496.1273629}) or specific optimization procedures designed to penalize changes in covariance structures (e.g., \cite{Angelosante_ICASSP2011}). 
Also, change point detection procedures in estimates of covariance dependencies can be used, e.g. \cite{DEGOOIJER20061892,GALEANO2014262,Yu_icassp2014_6854453,Gibberd_icassp2015_6854087}.
A more exhaustive and recent list of references can be found for instance in \cite{Gibberd2015,2016arXiv161007435H}.

The present work departs from these approaches because the full covariance structure is not required to stationary. Indeed, for interpretation in macroeconomics, only the global dominant structure needs to be constant and not all pair-wise correlations. Hence, we first extract a global structure by extracting features from the correlation-based graphs (as proposed in econophysics) before using  multivariate time series segmentations, leveraging here the methods from~\cite{Frecon2014}.

\subsubsection{Financial Integration in classical Macroeconomics}
For studying international financial integration, finding relevant eras of stable financial integration is an important  task, yet one that is difficult.
For instance, it was shown in \cite{bekaert1998capital,bekaert_dating_2002,bekaert2011segments} that critical thresholds between isolation and integration may be crossed several years after the facts causing a change of integration.
In  \cite{o1999globalization,obstfeld2004global}, international financial integration is assessed using price and volume indicators at the country level and based on the search for a ``U-shaped curve": The first era of high integration runs from the Industrial Revolution to the First World War, and the second era begins in the early 1980s.  The bottom of this U-shaped curve lies from the Great Depression to the 1950s. 
An issue in such analysis is precisely to determine the extent to which all countries converge \cite{lucas1990doesn,williamson2007global}, as well as to fix the thresholds between segmentation and integration (cf. e.g., \cite{bekaert_dating_2002} and \cite{bekaert2011segments}). 

On a theoretical level, two possible hypotheses are accepted to build a relevant periodization of the history of international financial integration. The first hypothesis, which is the most commonly accepted, is that of a periodization based mainly on the {\it institutional framework} of international capital markets. The alternative hypothesis is a periodization based on {\it private decisions} (or market decision if one thinks about the aggregate effects created by private decisions in front of the market), leading to massive changes in the pattern of international capital flows and therefore to networks' structural changes.

\paragraph{Institutional framework hypothesis}
The history of international monetary relations proposes a canonical chronology of the stages of international financial integration after World War II. This well-established periodization involves 3 major eras during our study period: the break-up of the so-called Bretton-Woods fixed exchange system in 1971; a period of turbulence between 1971 and 1975; then the affirmation, with the Kingston agreements and the introduction of flexible exchange rates in 1976, of a new era in international monetary and financial relations.

\paragraph{Private decisions hypothesis}
Alternatively, the history of capital movements identifies four distinct periods: a first period until the end of the 1960s when capital movements are dominated by a North-North pattern, a second period until the beginning of the 1980s with massive North-South capital flows, a third period until the middle of the 1990s with a return to a North-North pattern, and a fourth period of new gradual and selective rise of North-South capital flows with the advent of emerging markets. The different periods are mainly determined by the growth differential between advanced and developing economies. Additional factors in these four periods are:
i) international lending / borrowing activities in foreign currency, particularly in dollars, which are growing (+ 380\% between 1973 and 1980); 
ii) oil shocks in the 1970s which exacerbate excess liquidity, and hence the need to lend of international banks, but also the need to borrow of non-oil developing countries; 
iii) awareness of the unbearable nature of the accumulated indebtedness which reduces the North-South capital flows after 1980,
and iv) the long-lasting slowdown caused by the Global financial crisis of 2008 that increases foreign direct investments in developing (especially emerging) economies, which exceed those to advanced economies. 

\paragraph{Summary}
The economist is then confronted to various possible interpretations and periodization of financial integration.
The assessment of  what are relevant eras of stable global financial integration, in historical perspective, is thus
a challenging task that can be complemented by a data-driven segmentation, as is proposed here.
We show here that it is possible to escape arbitrary decisions by looking globally at the structure of dependences between price indicators.

\begin{figure*}[t]
%\centerline{
\includegraphics[width=.38\textwidth]{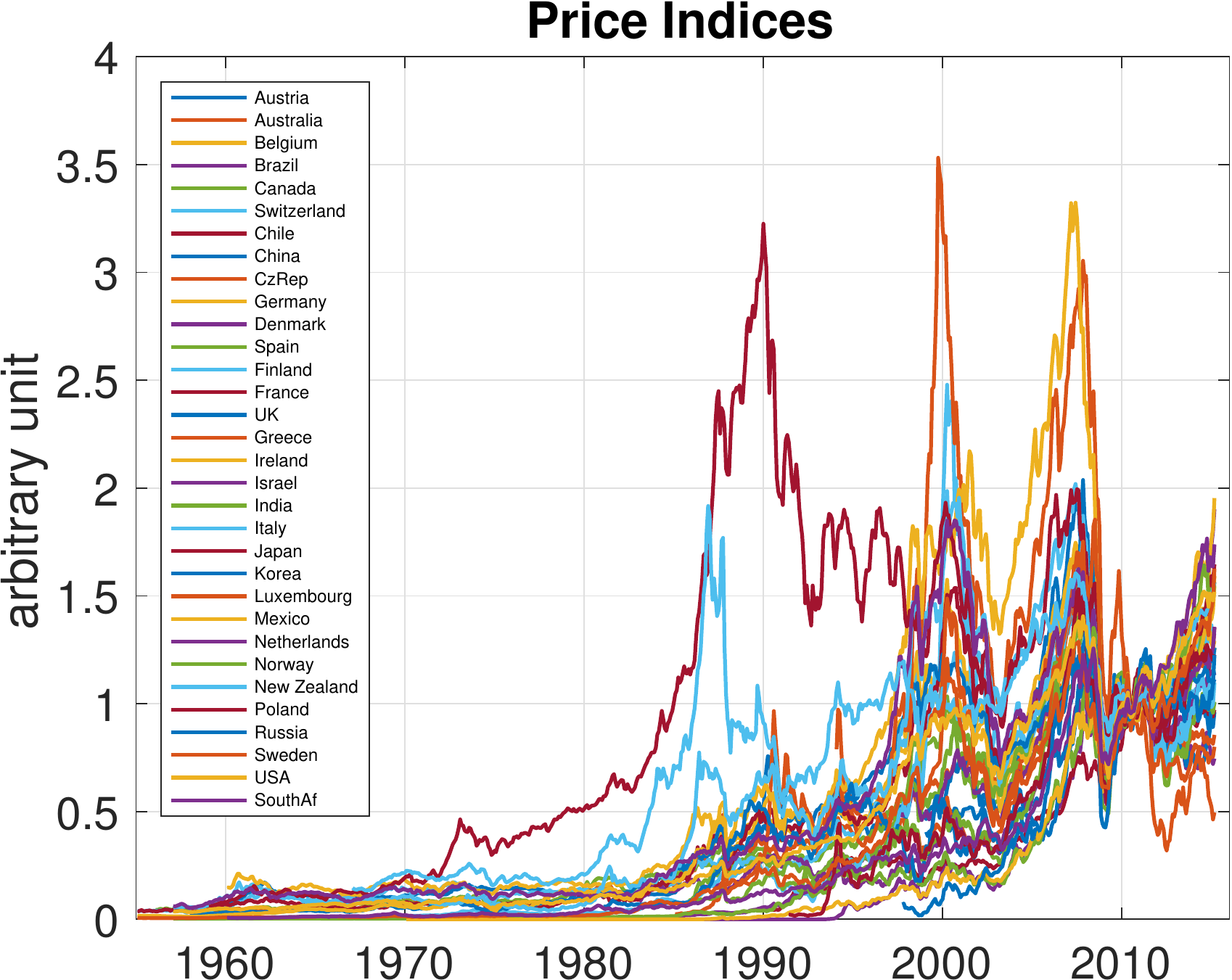}
\begin{minipage}[b]{.22\textwidth}
\begin{center}
\includegraphics[width=.9\textwidth]{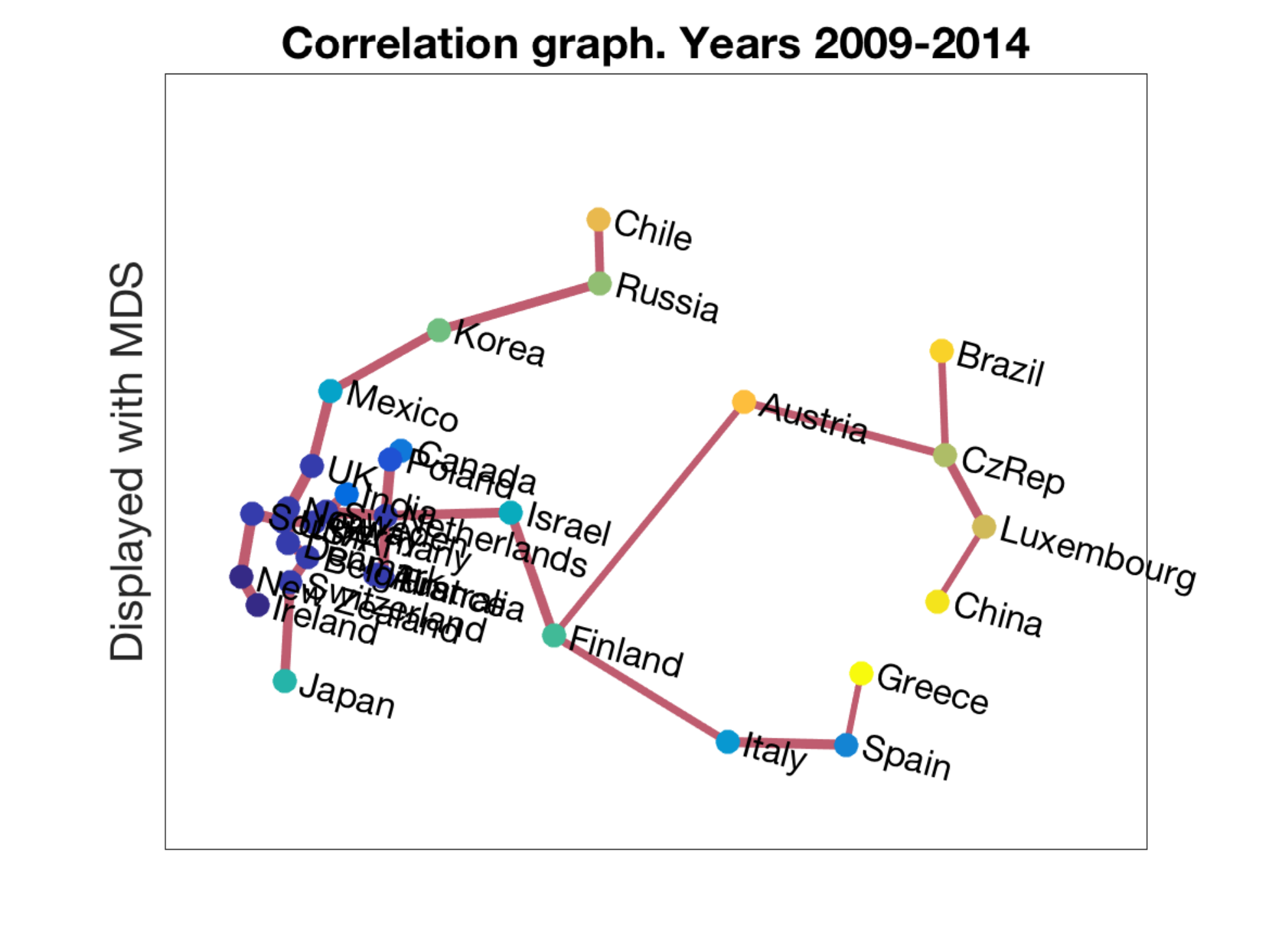} 
\includegraphics[width=.9\textwidth]{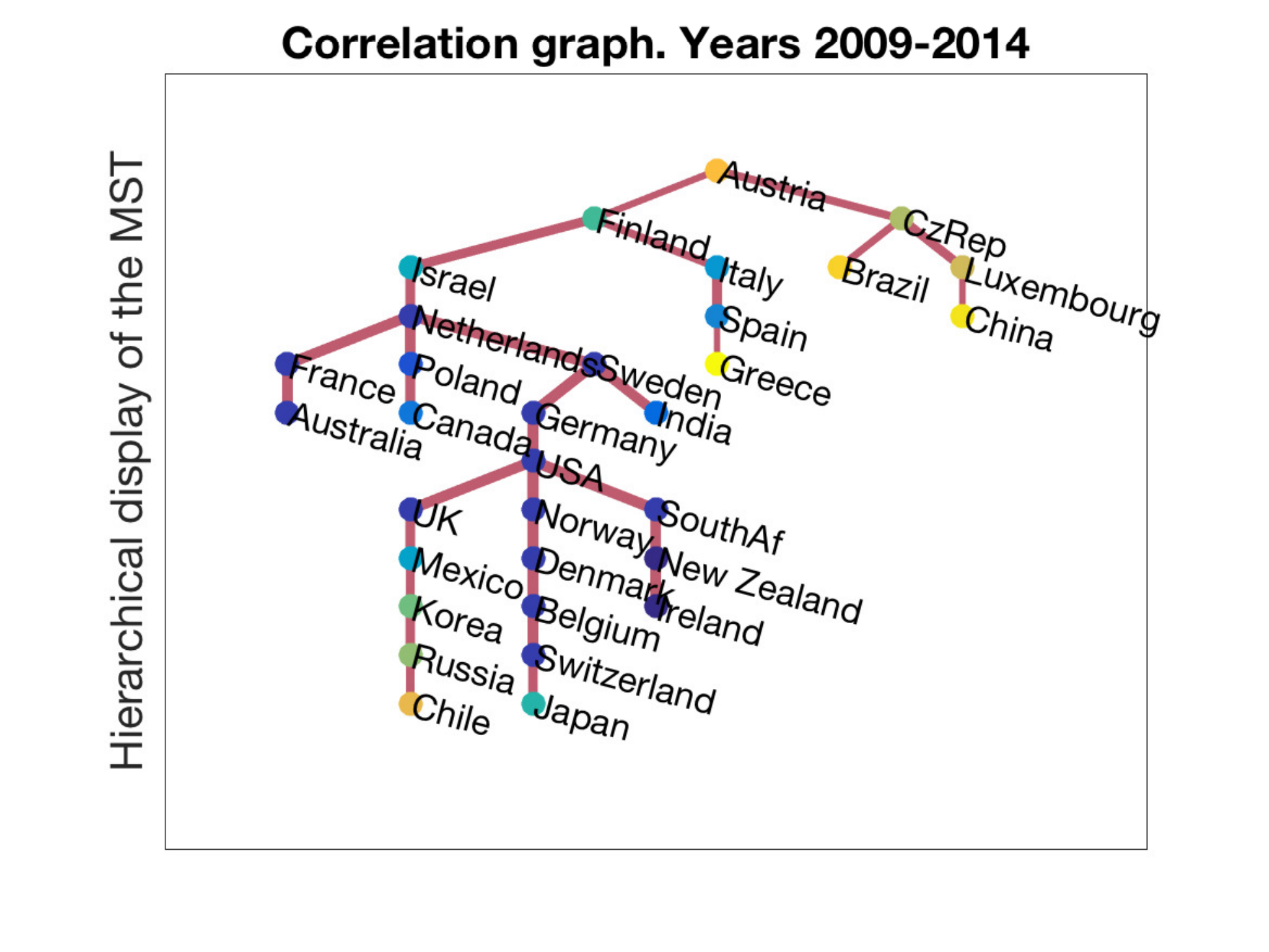} 
\end{center}
\end{minipage}
% Remplacer par MDS plane 2010-2015 ? Disposition des noeuds pas intuitive sur celui-là comme vu ensemble  
\includegraphics[width=.38\textwidth]{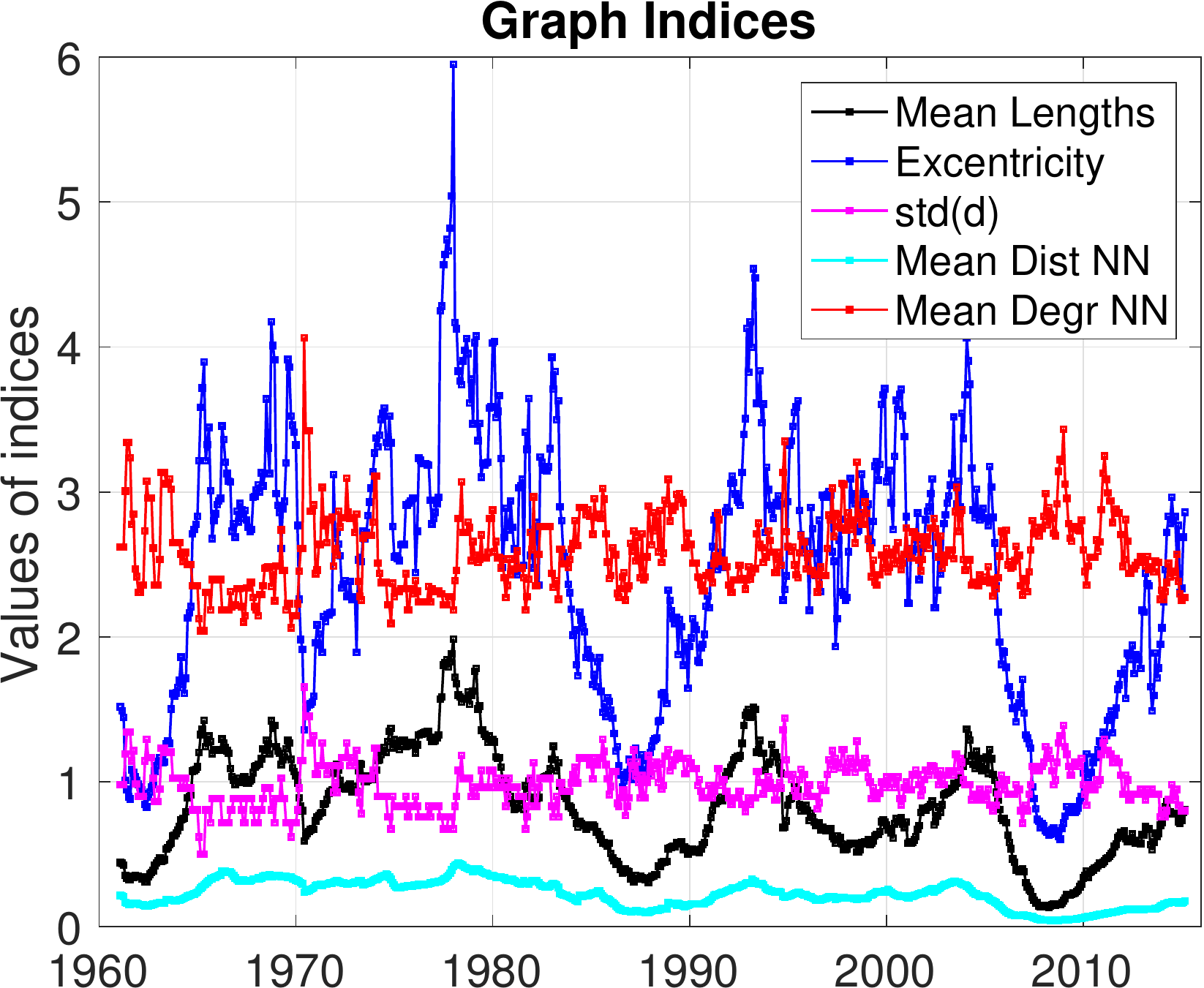}

\caption{\label{fig:figa} {\bf Analysis procedure: 3-steps.} Step 1 (left):  Price index times series; 
Step 2 (middle) : Covariance estimation, represented here by its Minimum Spanning Tree that is used to describe the overall dependence structure. Top: the MST is displayed using a representation with classical MDS and bottom: using a hierarchical view of the tree. 
Step 3 (right):  Time series of the evolution of the graph-based dependence structure, in 5 connectivity and distance indices, on which are used temporal segmentation (Note: the year used as label is set at window end).}
%\caption{\label{fig:figa} {\bf Econometrics.} (a) Price index times series (b) MDS (or MDS) representation (c)  Graph based index time series}
\end{figure*}

\subsection{Contributions and outline} 

%In this context, the present work aims to achieve an endogenous determination of periods, or eras,  delineated by critical changes in the structure of global equity markets. 
%Assessing local stationarity directly on the sliding estimates of the covariance structure would result into a stringent procedure requesting that correlations between \emph{all pairs} of countries remain constant along time. 
%Elaborating on ideas in \cite{mantegna1999hierarchical,bastidon2016form}, a graph-based segmentation procedure is proposed, which puts the focus on the global and topological structure of the graph, rather than on pairwise relations. 

The contribution is to propose a procedure to study correlations between equity markets with the objective of finding 
relevant stable eras of financial integration.
The study is conducted over asset price time series like in most econometric literature (e.g. \cite{engle2002dynamic}). 
The dataset used is detailed in Section~\ref{sec:data}): it consists in a collection of national equity market indices from (up to) 32 different countries for the years 1955 to 2015.
%  analyzed in long sliding windows of several years (with overlaps). 

The procedure consists of three steps that are sketched in Fig.~\ref{fig:figa} and detailed in Section~\ref{sec:method}. 
The first step constructs, from the empirical estimation of the covariance structure, a graph of relations between countries, based on a minimal spanning tree (MST) from the correlation distance; it gives a time-dependent graph thanks to doing this analysis on (long) sliding windows of several years (with overlaps).
The second step computes a set of 5 topological indicators, quantifying the distances and connectivities in the constructed MST graphs along the time.
%(average distance to the nearest neighbors, average length of paths, eccentricity) 
%and connectivity
% (average degree of the nearest neighbors in level and standard deviation)
%which are associated with the minimal spanning trees constructed over rolling periods. 
It thus replaces the analysis of the 32 asset price times with that of 5 time series representing the global covariance graph structure evolution along time. 
In a third step (illustrated in~Fig. \ref{fig:figb}), a multivariate piece-wise constant signal denoising procedure, based on functional optimization \cite{Frecon2014}, is applied to these topological index time series to detect change points in the structure of the graph. The whole procedure allows to identify globalization eras characterized by locally stationary network structures. Our results are presented and discussed in relation to macroeconomic data in Section~\ref{sec:res}.

% This study precisely aims to address the question of financial integration eras segmentation raised by the macrofinancial literature, using network indicators derived from topological representations of assets portfolios.  
% 1 - As regards the issue we are departing from two distinct reference literatures: one in financial macroeconomics and one in econo-physics. 
% 2 - denoising de series temporelles  d economie ou finance en vue de segmenter 
% 3 - one could have done the following : estimate cov in sliding window and a posteriori try to segment into era where cov matrix remain constant ? 
% why not ? less original (eco), classical local stationarity does not permit to compute global connectivity indices, would be more stringent as requires a that correlation between specific pairs of country remain constant, where as here only interested in global structure of the graph that remains constant, here want to use global indices of graph as a way to interprete a posteriori; 

% Ref eco sur approche par corr/cov moins originale

% Ajout
% dans le détail des metadata la construction des indices n'est pas totalement homogène sur deux points. 1. Couverture totale des entreprises côtées ou limitée à plusieurs centaines de titres. 2. Introduction dans certains cas de moyennes pondérées remplaçant les moyennes arithmétiques sur la fin de période (vers 2000) quand les indices pondérés se généralisent
% on peut alléger la présentation de la base de données

\section{Data}
\label{sec:data}

\subsection{Dataset.}  The dataset used comprises equity market price indices for up to 32 countries world-wide, compiled by the Monthly Monetary and Financial Statistics of the OECD,
%Organization for Economic Co-operation and Development, 
a reference institution for economic history data.
These indices are representative at national levels, including the closing daily values of all listed shares or a broad coverage, normally expressed as simple arithmetic averages. 
Data are collected monthly, from Jan.~1955 to 2015 and available at the FRED database (\url{<https://fred.stlouisfed.org/>}). 
In the fist years, only 13 countries are documented, the maximum number of 32 being attained from 2005 on. We have consistently checked that, {with the exception of the 1960s where there is a jump from 13 to 18 countries}, there is no correlation between this increase and the evolutions reported hereafter. 

\subsection{ Multivariate time series.}  Choice has been made here to work with price indices in level, as illustrated in Fig.~\ref{fig:figa} (left).  
Indeed, the standard economic theory of assets prices states that all agents in equity markets have rational expectations and that markets follow random walks around trends \cite{samuelson1965proof}. 
A complementary literature further states that there is also a second category of agents whose expectations are not (or only partially) rational and thus that asset prices are affected by additional noises \cite{shleifer1990noise} or bubble mechanisms, i.e., \emph{temporary deviations of asset prices from fundamental values}, due, for example, to liquidity trading or to waves of optimism or pessimism \cite{bernanke2000monetary}. 
Thus the correlations matrices derived from long term price series in level describe the economic fundamental dimension of the integration of global equity markets, rather than its purely speculative dimension.

\begin{figure*}[h]
\centerline{
\hfill (a) \includegraphics[width=.46\textwidth]{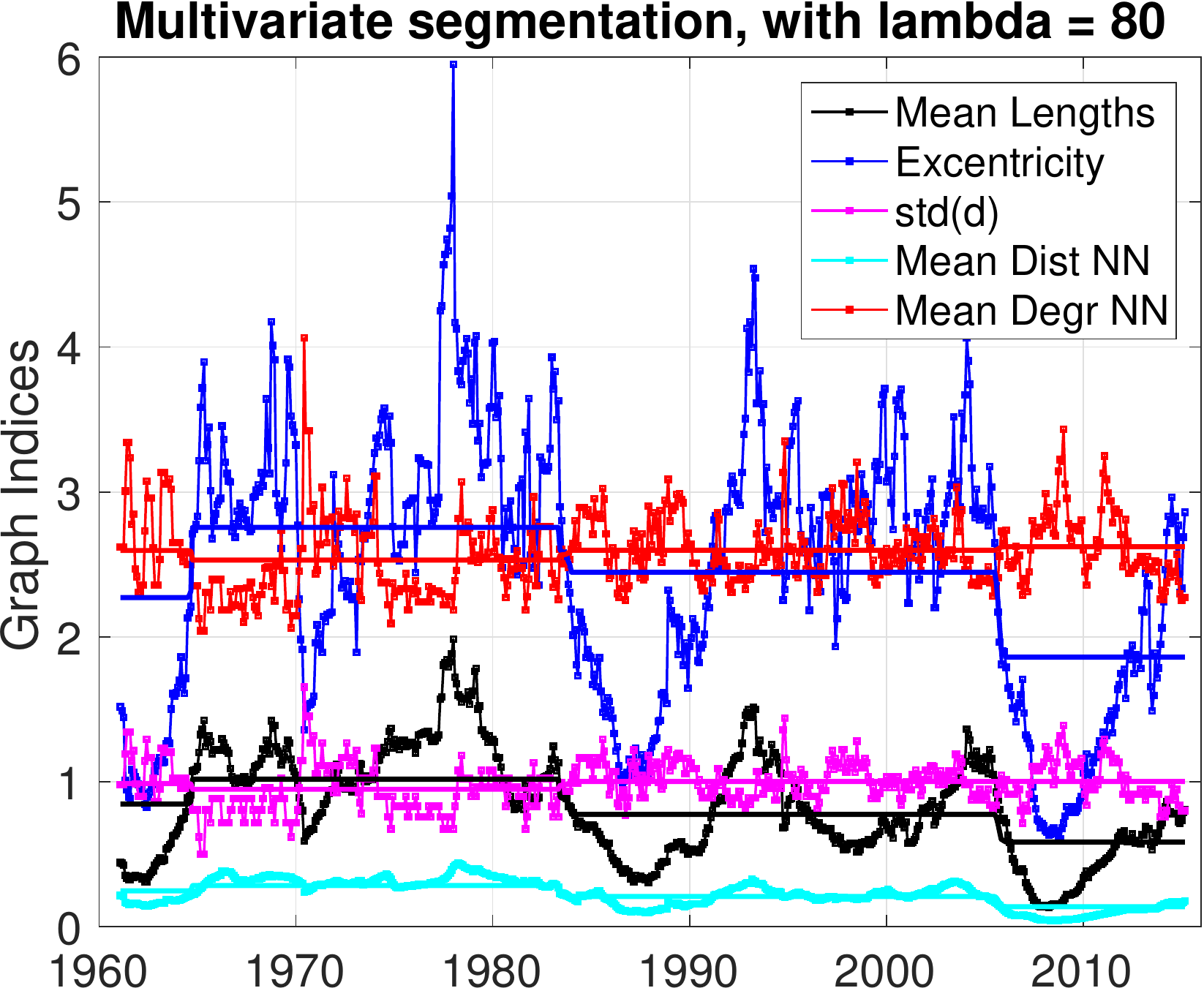} \hfill
(c) \includegraphics[width=.46\textwidth]{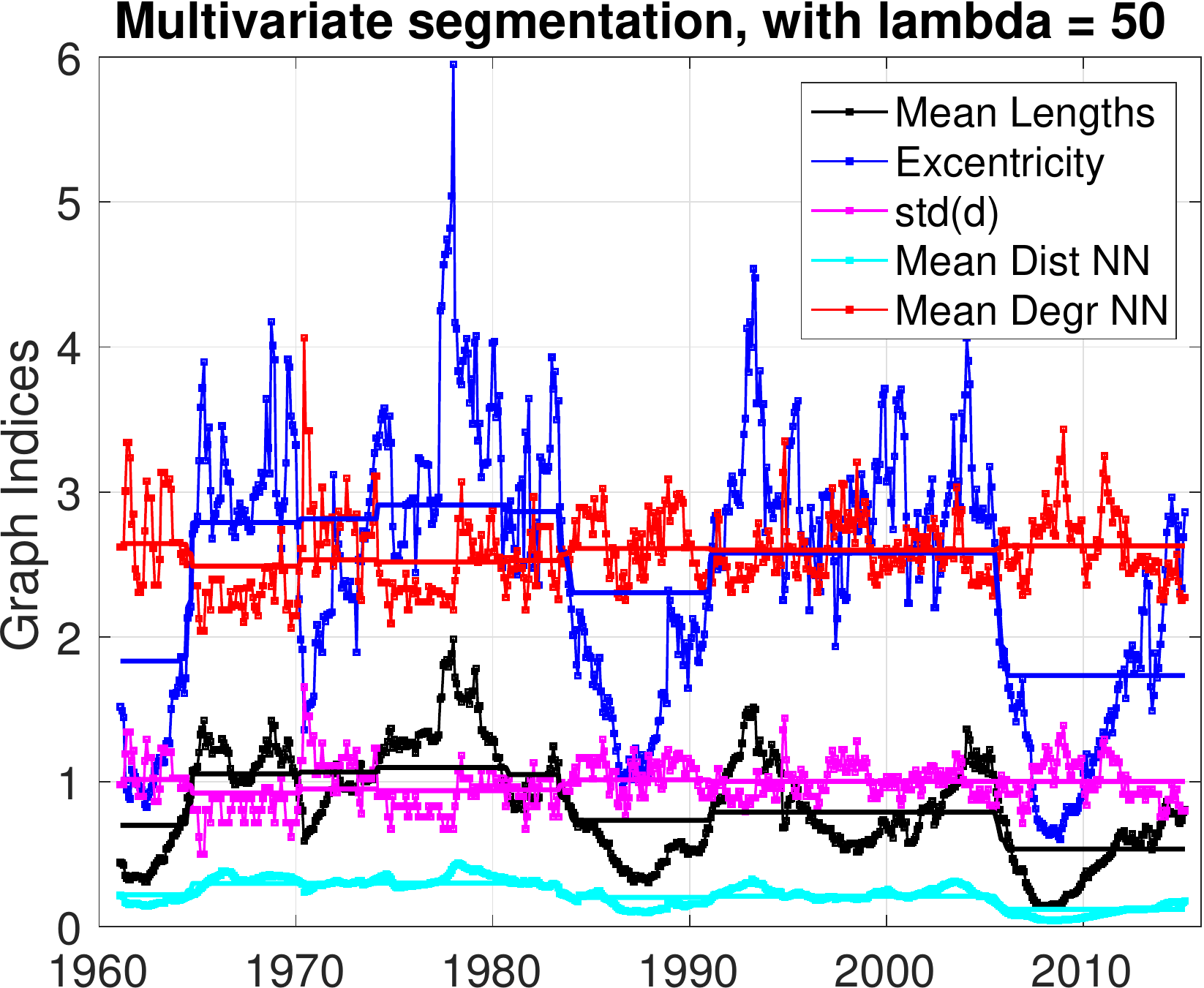} \hfill
}
\centerline{
\hfill (b) \includegraphics[width=.46\textwidth]{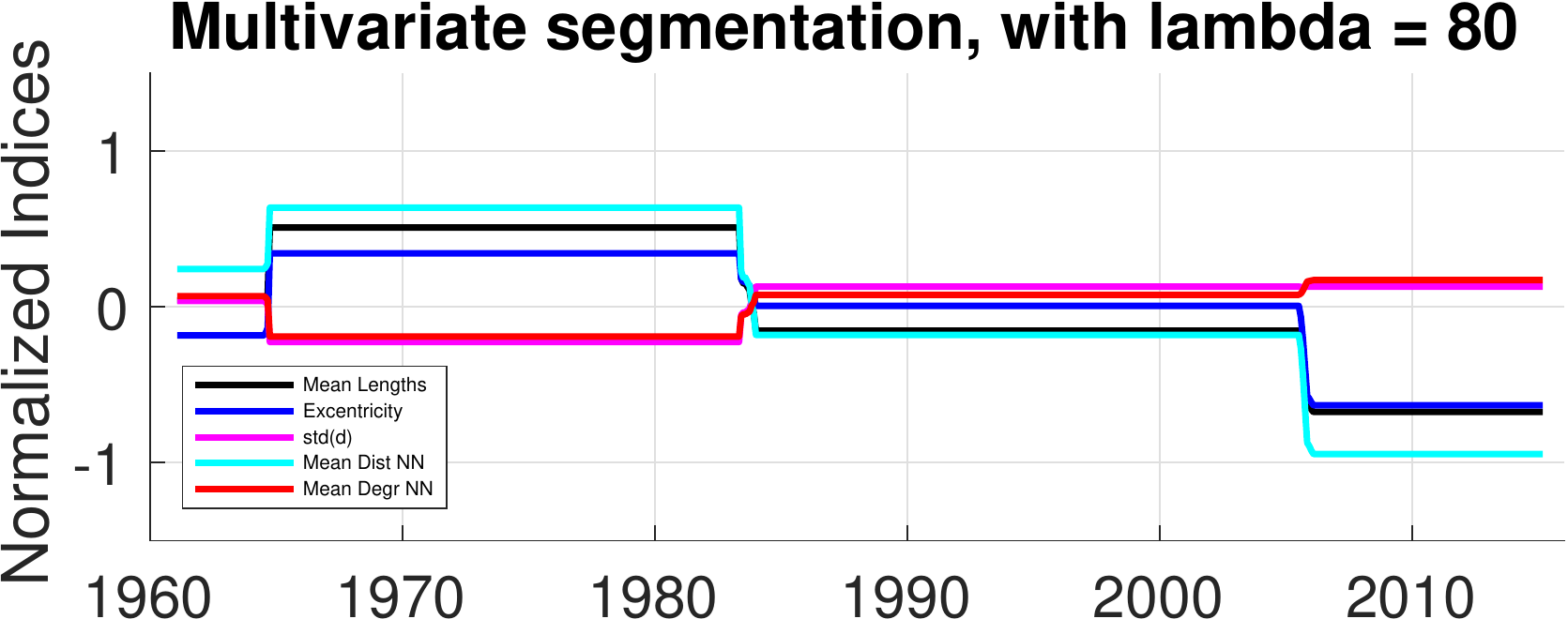} \hfill 
(d) \includegraphics[width=.46\textwidth]{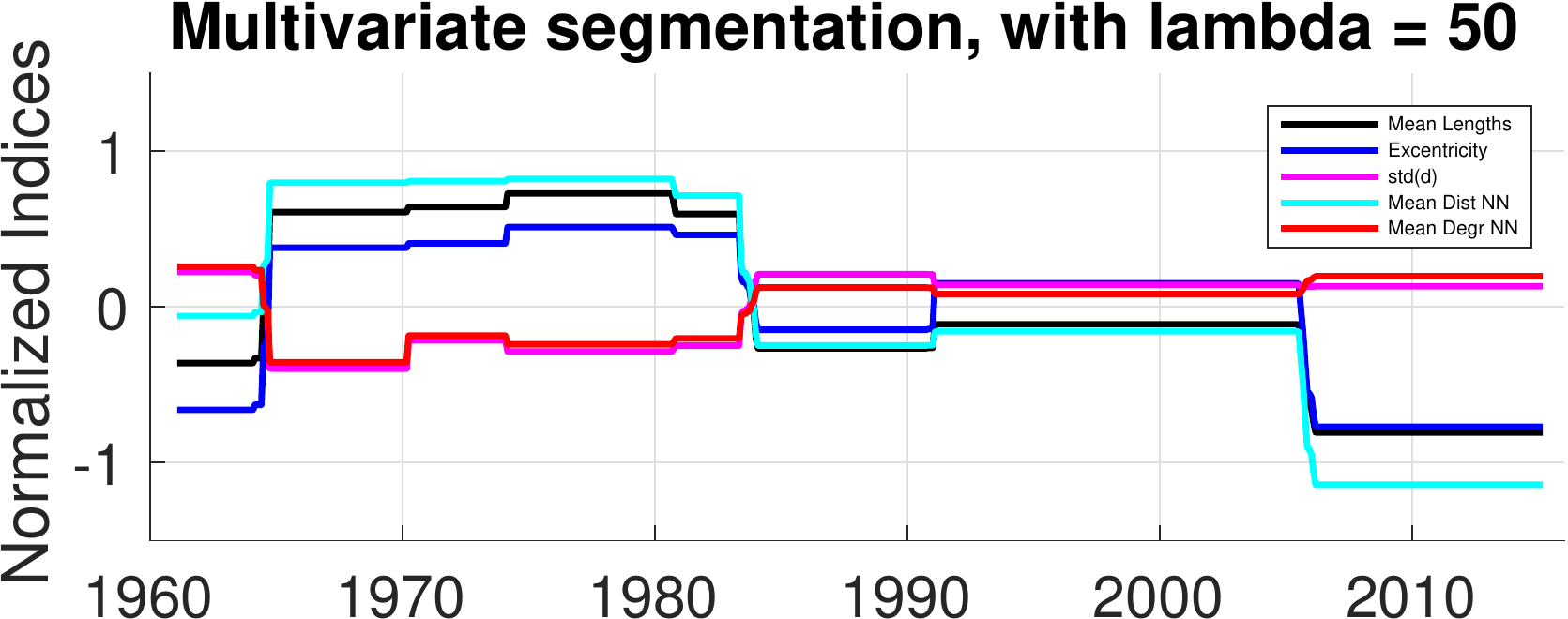} \hfill
}
\caption{\label{fig:figb} {\bf Era segmentation from Multivariate times series.} (a) and (b) in 4 eras, (c) and (d) in 8 eras.
The top figures show the multivariate indices and the obtained piecewise constant levels. The bottom figures complement that by showing only the obtained piecewise constant levels of indices if centered and normalized, for a better comparison of their evolutions
(Note: the year used as label is set at window end).}
\end{figure*}

\section{Method: Graph-based segmentation in eras}
\label{sec:method}

\subsection{Step 1: Graph construction.} 
%Assessing financial integration directly from the multivariate asset price time series or from their estimated covariance structure would prove challenging. 
Instead of using directly the multivariate asset price time series or their covariance, 
it has been chosen to rely on the topological graph structure that accounts for the global dependence structure of 
multivariate data~\cite{mantegna1999hierarchical}. To that end, from the signed pairwise correlations $\rho_{ij}$ between two countries $i$ and $j$ in a sliding window, the distance $d_{ij} = 1-\rho_{ij}^2$ is computed and then used to construct a Minimal  Spanning Tree (MST) on the covariance structure.
This MST offers a global topological representation of the covariance, as was originally proposed in econophysics for portfolio studies~\cite{mantegna1999hierarchical}. It is associated with the subdominant ultrametric, providing intuitive economic interpretations because of the tree structure.
%Instead, it has been chosen here to rely on the assessment of topological graph structure that accounts for the global dependence structure of the multivariate data rather than on the details of their pairwise correlations. 
%To that end, covariance is estimated in a sliding window and an affinity matrix is constructed from unsigned {\bf or signed ??} correlations.
%A minimal minimal spanning tree (MST) is then constructed, as it is one  of the most commonly used topological representation in the econophysics portfolio literature, both because it is associated with the subdominant ultrametric and because subtrees usually correspond to intuitive economic interpretations \cite{mantegna1999hierarchical}. 
An independent MST is obtained for each sliding window. 

\subsection{Step 2: Graph Indices.} To characterize the structure of the graphs, features or indices are computed from the MST, quantifying its topology.
To quantify distances, three indices are used. 
The mean distance to the nearest neighbors (NN) is the most basic measure.
The mean length of paths (from each node to all the other nodes of the network) is a most sophisticated distance measure. 
The eccentricity (i.e.,  maximal path length) is a convenient way to display how the less integrated nodes converge.
To measure connectivity, two indices are added: 
the standard deviation of degrees, and the mean of the degree of the nearest neighbors in the MST.
Both indices are expected to increase when the structure of the network changes from a linear to a star network. 
All together, these indices describe if a graph is large, and if it looks more like a star or a linear structure.
% We are using two types of network indicators, both computed from the minimal spanning trees: distance measures, and connectivity measures. . . \\
The complete structure can also be examined by looking at a representation of the MST (see Fig. \ref{fig:figa} (b)).

\subsection{Step 3: Multivariate joint piece-wise constant denoising.} The denoising of the time series indices $\mathbf{X}  = \{X_k(t), k=1,\ldots,K, t=1\ldots, T \}$ is achieved by minimizing a functional form consisting of a data fidelity term balanced by a penalization term aiming to suppress spurious changes while preserving actual ones:
\begin{equation}
\label{eq:l1denoising}
\displaystyle \mathbf{Y}(t) = \makebox{arg}\min_{\mathbf{U}} || \mathbf{X} - \mathbf{U}||^2 + \lambda \sum_t \sqrt{ (\sum_k |U_k(t+1)-U_k(t)|)^2}.
\end{equation}  
The use of total variation (or $l^1$-norm) in the penalization term favors piece-wise constant solution $\mathbf{Y}(t)$. 
Its combination with a $l^2$ norm across components permits to favor the detection of change-point jointly across components, i.e., at the same time location. 
The parameter $ \lambda $ mitigates the number of change points actually produced. 
Intuitively, this mixed-norm penalization procedure performs a non-linear smoothing of the data, ideally smoothing when no change occur hence denoising data, while preserving actual changes. 
The minimization of that convex but non differentiable functional is achieved by means of proximal based primal-dual algorithms. 
For more details on the employed algorithm, readers are referred to \cite{Frecon2014,Condat2013}. 

\subsection{Visualisation of financial networks.}
\begin{figure*}[h]
(a)
\centerline{\includegraphics[width=.4\textwidth]{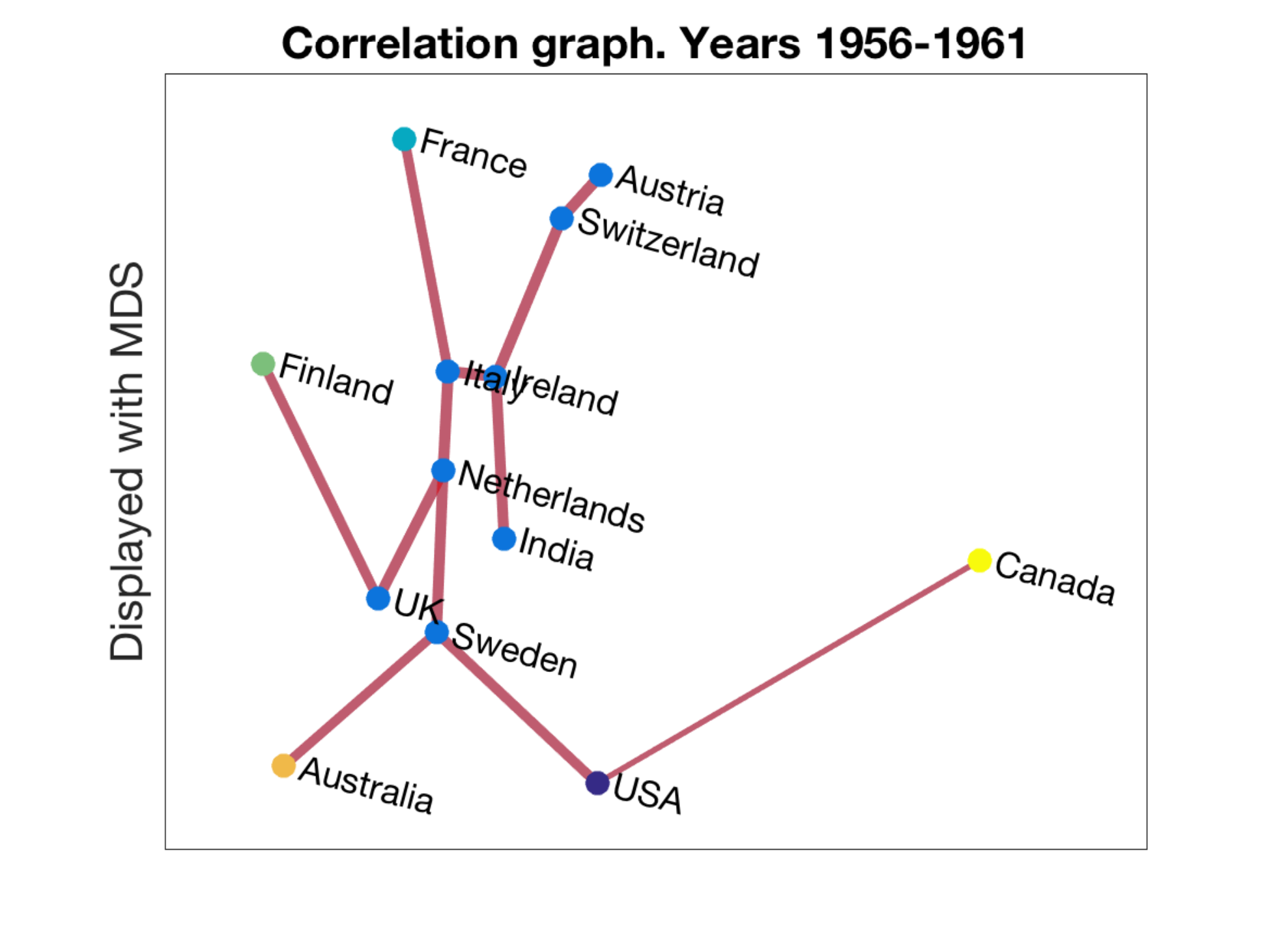}
\includegraphics[width=.4\textwidth]{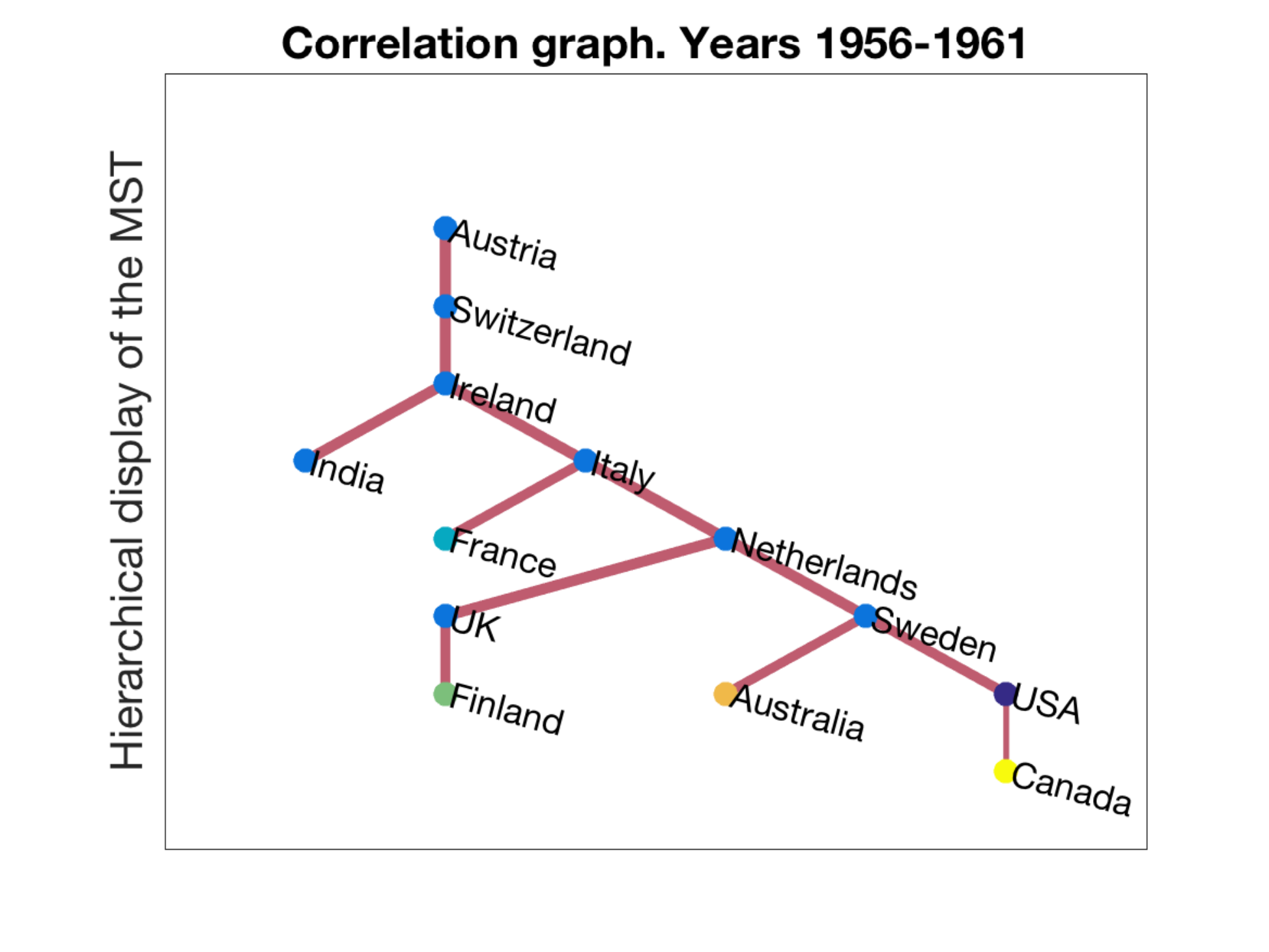} }

(b)
\centerline{ \includegraphics[width=.4\textwidth]{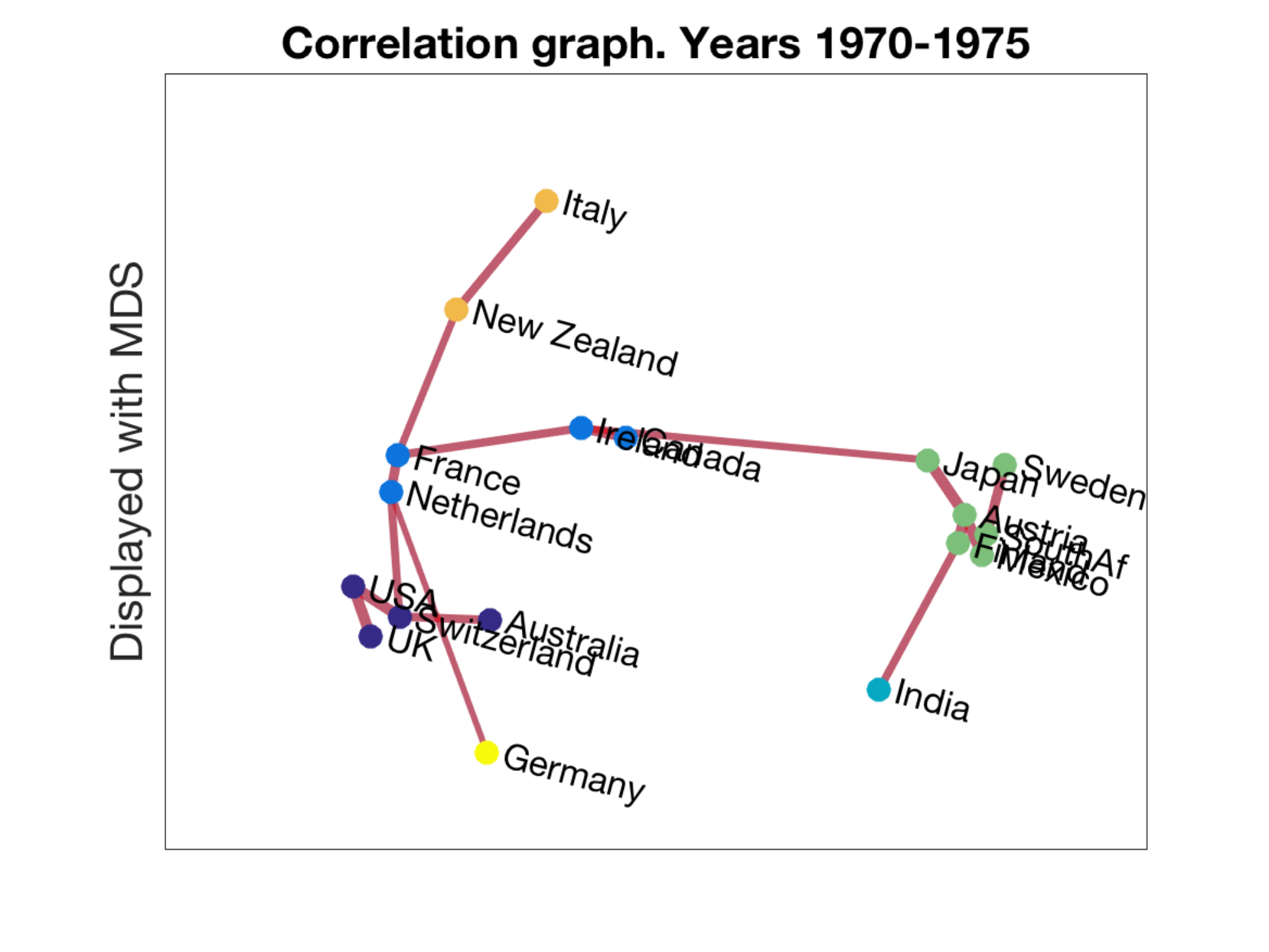}
\includegraphics[width=.4\textwidth]{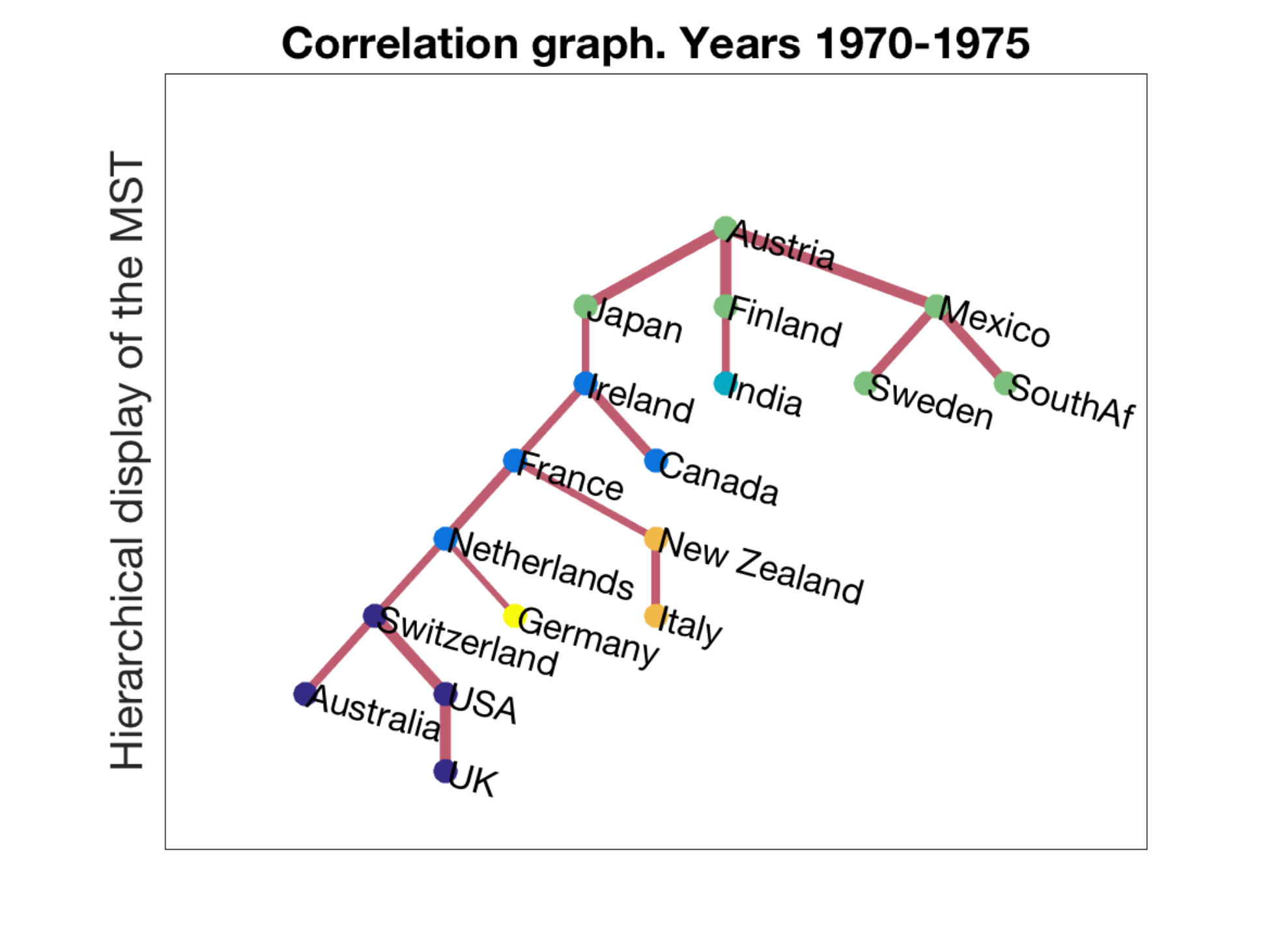} }

(c)
\centerline{ \includegraphics[width=.4\textwidth]{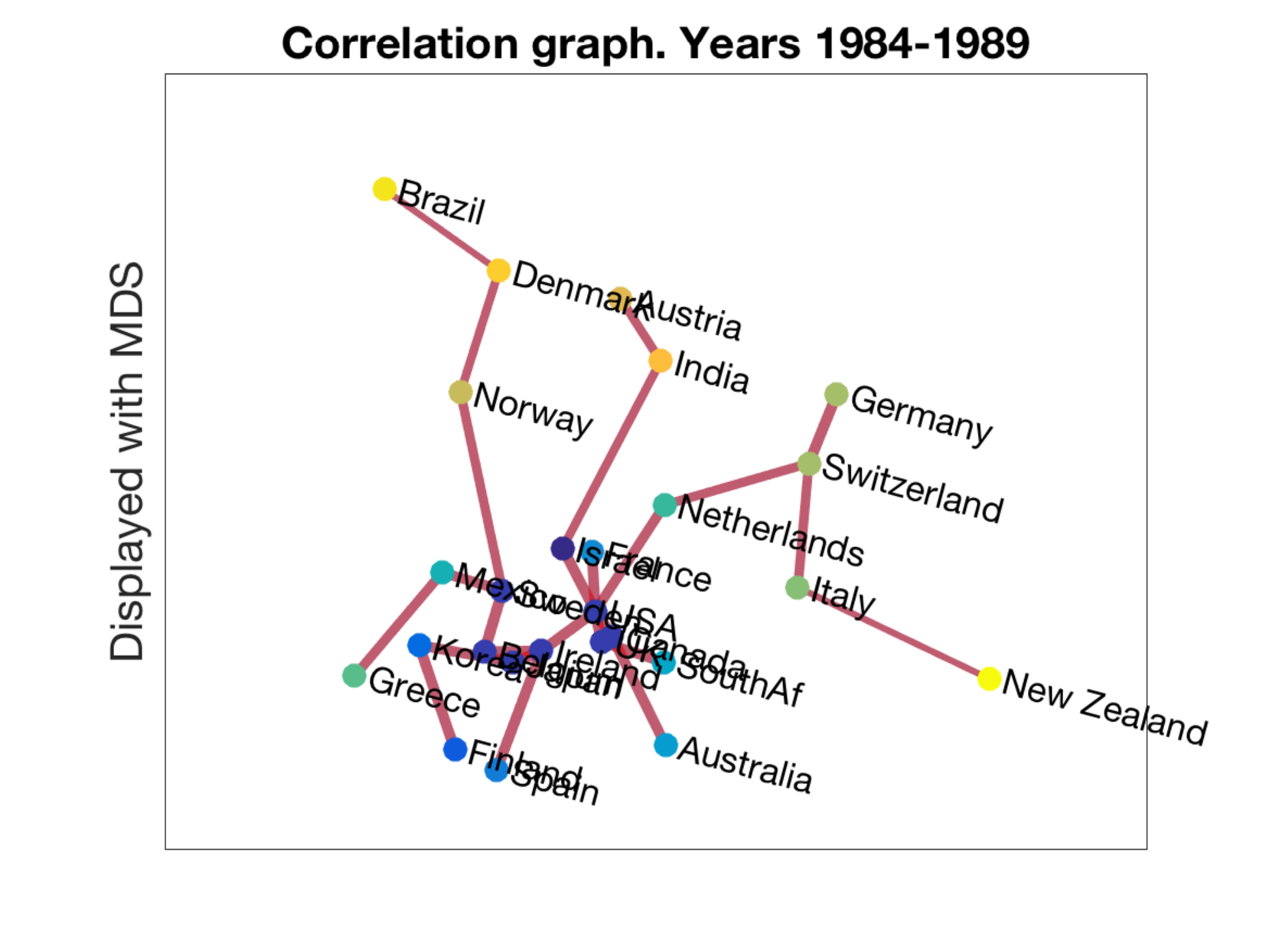}
\includegraphics[width=.4\textwidth]{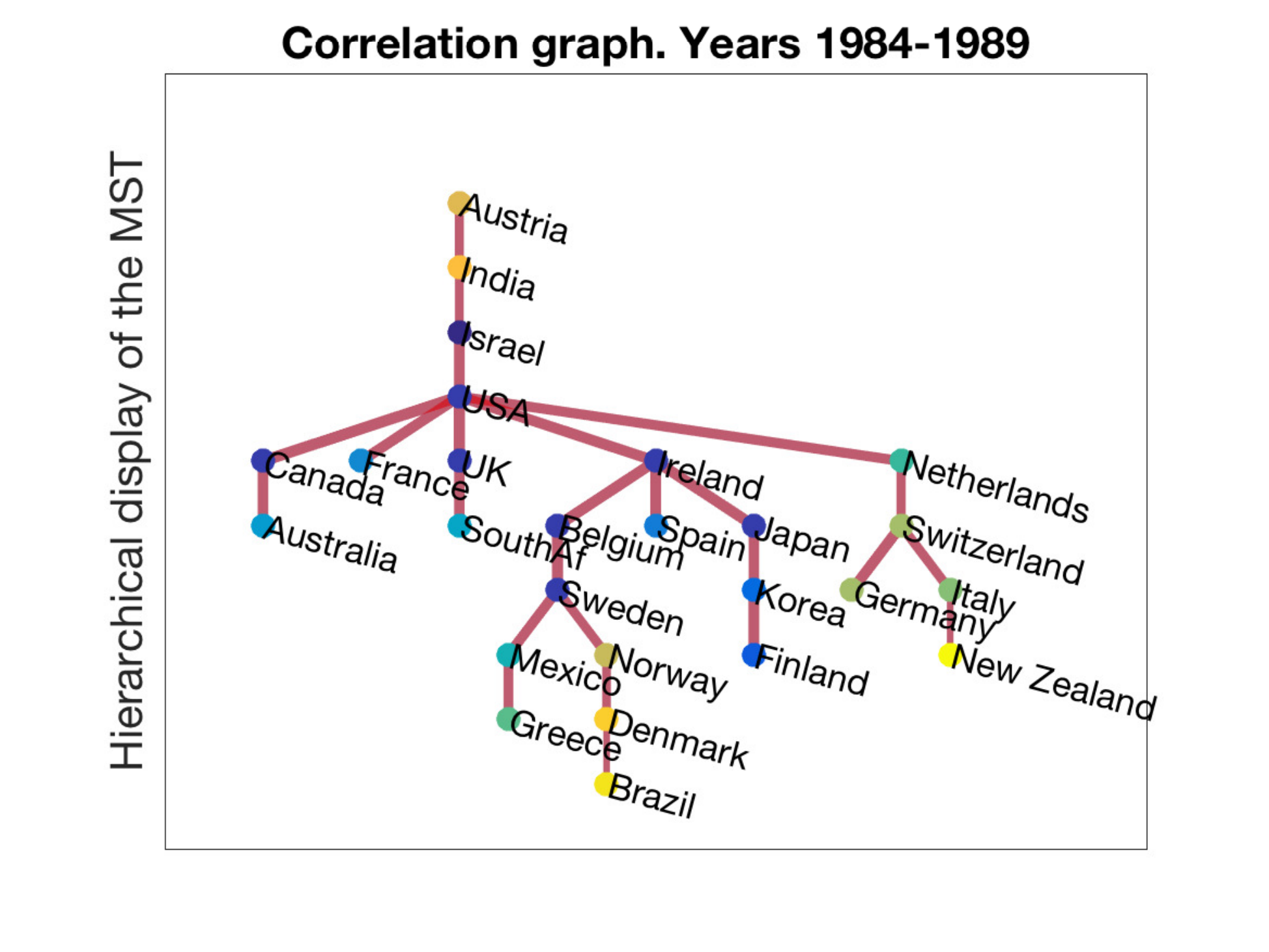} }

(d)
\centerline{ \includegraphics[width=.4\textwidth]{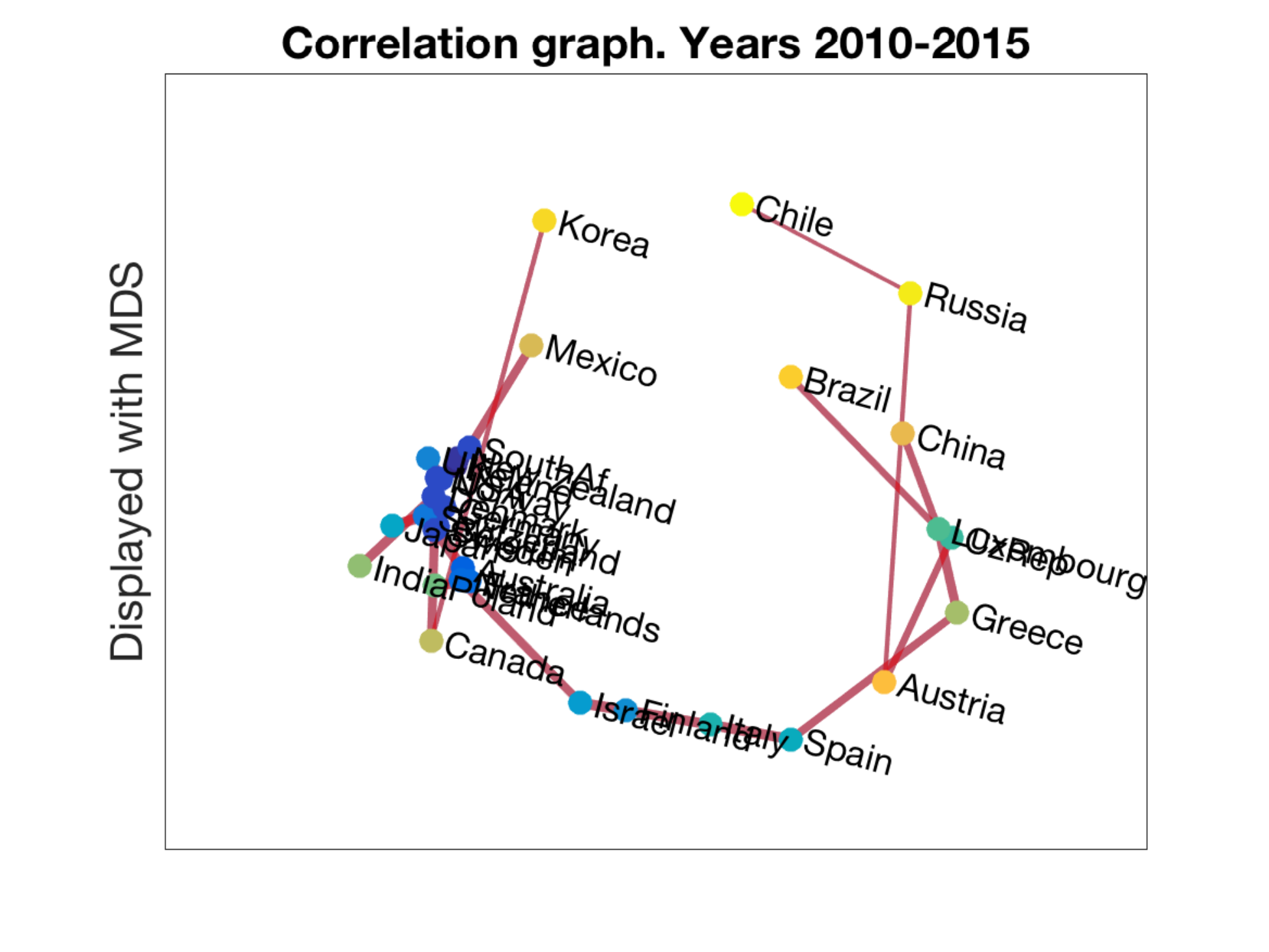}
\includegraphics[width=.4\textwidth]{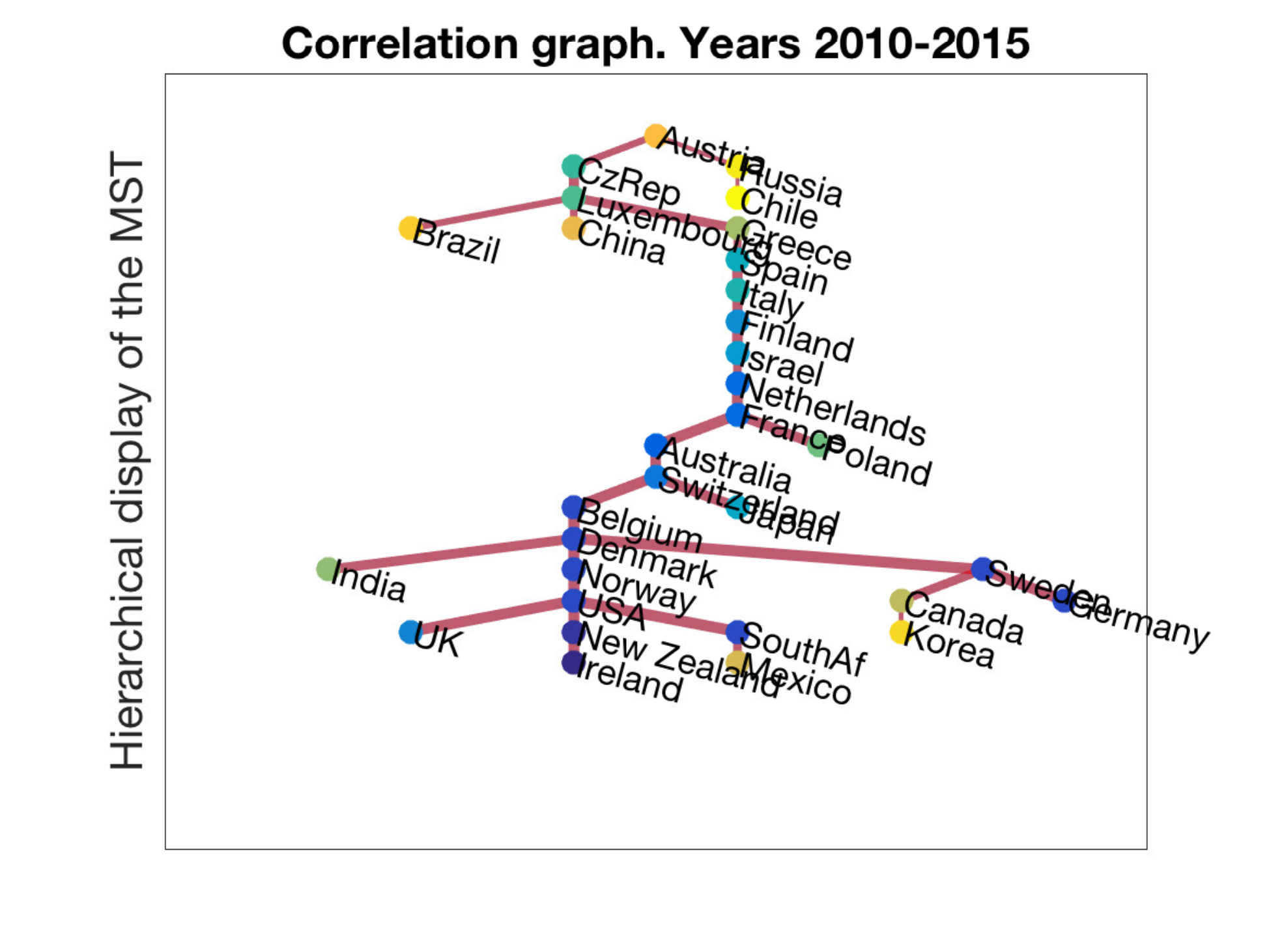} }
%\centerline{\hfill (a) \hfill (b) \hfill}
%\centerline{
%\includegraphics[width=.4\textwidth]{fig3_1956_tree_MDSplane}
%\includegraphics[width=.4\textwidth]{fig3_1970_tree_MDSplane}
%%\includegraphics[width=.5\textwidth]{fig3_1984_tree_MDSplane}
%%\includegraphics[width=.5\textwidth]{fig3_2010_tree_MDSplane}
%}
%\centerline{
%\includegraphics[width=.4\textwidth]{fig3_1956_tree_MSTplane}
%\includegraphics[width=.4\textwidth]{fig3_1970_tree_MSTplane}
%%\includegraphics[width=.5\textwidth]{fig3_1984_tree_MSTplane}
%%\includegraphics[width=.5\textwidth]{fig3_2010_tree_MSTplane}
%}
%\centerline{\hfill (c) \hfill (d) \hfill}
%\centerline{
%%\includegraphics[width=.5\textwidth]{fig3_1956_tree_MDSplane}
%%\includegraphics[width=.5\textwidth]{fig3_1970_tree_MDSplane}
%\includegraphics[width=.4\textwidth]{fig3_1984_tree_MDSplane}
%\includegraphics[width=.4\textwidth]{fig3_2010_tree_MDSplane}
%}
%\centerline{
%%\includegraphics[width=.5\textwidth]{fig3_1956_tree_MSTplane}
%%\includegraphics[width=.5\textwidth]{fig3_1970_tree_MSTplane}
%\includegraphics[width=.4\textwidth]{fig3_1984_tree_MSTplane}
%\includegraphics[width=.4\textwidth]{fig3_2010_tree_MSTplane}
%}
\caption{\label{fig:figc} {\bf Visualization of the financial networks} in four windows (6-year length), each situated
inside a specific era of the obtained segmentation: (a) 1956-1961, (b) 1970-1975, (c) 1984-1989, (d)  2010-2015.
The left figures are displayed using MDS and the right ones using hierarchical view of the tree (see end of Section~\ref{sec:method}).
\vspace*{-3mm}}
\end{figure*}

For visualisation and inspection of the results, we display in Fig. \ref{fig:figa} (b) and later on in Fig. \ref{fig:figc} the MST with two different techniques: first, we use classical MDS (Multi-Dimensional Scaling) \cite{BorgGroenen2005} of the correlation distance $d_{ij}$, to embed in two dimensions the nodes of the MST. This visualization keeps all the covariance structure and reveals the distances behind the financial integration measured here. Second, we show a hierarchical view of the MST as a tree, to emphasize the relative positions of the countries. This highlights the opposition between the tree/star structure of the MST. The colors used on the nodes are derived from a hierarchical clustering using a complete linkage algorithm (as in \cite{bastidon2016form}). We do not use this clustering for interpretation in the present work, as it mostly shows that classes group together neighboring nodes in the MST, as expected. The widths of the graph edges are proportional to the unsigned correlation.

%\begin{figure*}[h]
%\centerline{
%\includegraphics[width=.33\textwidth]{fig3_1970_tree_MSTplane}
%\includegraphics[width=.33\textwidth]{fig3_1984_tree_MSTplane}
%\includegraphics[width=.33\textwidth]{fig1b_tree_MSTplane}
%}
%\caption{\label{fig:figb} {\bf  }}
%\end{figure*}

\section{Results and discussions}
\label{sec:res}

\subsection{Parameter setting.}
Analysis are conducted in 6-year long sliding window, chosen as a trade-off: 
Long enough windows ensure estimation of covariance with sufficient quality, while short windows allow for a more precise detection of change. For countries whose data is not available from the beginning, correlation is computed if data exist for at least half of the window.  
%{\bf say something about the fact that at the begining not all countries are available, explain how this is delt with  ? missing data and covariance estimation. Should this go into the data section? }
For the optimization involved in Step 3, no attempt to automatically tune $\lambda$ to an optimal value has been used. 
Rather, $\lambda$ has been manually adjusted to match the intuition that over the 60 years spanned by data, there should be, by nature or even definition, only a few era of stable financial integration, ranging from at least $2$ up to, say, $10$. 
Still, neither the exact number of eras nor the location of the change points are chosen a priori. 
They constitute the outputs of the 3-step analysis procedure.  

\subsection{A 4-era segmentation.} A first segmentation, achieved for a large value $\lambda = {80}$, yields a 4-era segmentation in the time evolution of the structures of the graphs, as illustrated in Fig. \ref{fig:figb} (a) and (b), with change points located around years: 1964, 1983 and 2007. 

The change point located in 1983 is caused by a joint decrease in distance indices and increase in connectivity ones,  cf. Fig. \ref{fig:figb} (b). 
From a historical economics perspective, the end of 1982 corresponds to the onset of the developing countries debt crisis. 
This crisis is recognized as one of the largest financial crises of the 20th century,
% combining regulatory factors and prices shocks, 
resulting from the unprecedented increase in international capital flows along with the historical rise in assets prices volatility in the 1970s. 
As a consequence, North-North capital flows became dominant for the next three decades, while large North-South flows were observed before 1983. 
This is consistent with the lower decrease in the eccentricity of graph after the first change point, in comparison with the other distance indices. 
This suggests that the convergence of developing economies is slow during the 1982-2007 subperiod. 
It is also consistent with the USA becoming central in MST observed after 1983 (see Fig. \ref{fig:figc} (c)), having then 6 edges,
as opposed to the structure of the graph observed before 1983 (Fig. \ref{fig:figc} (a) and (b)) where it has only 2 edges and is less central, especially before 1964.
Globally, before 1983, the MST in the MDS plane (Fig. \ref{fig:figc} (a) and (b) top) show a core of advanced economies showing moderate integration that becomes, after 1983 (Fig. \ref{fig:figc} (c) and (d)) much more integrated in a compact core, as compared to developing or emerging economies. {Indeed the distribution of the distance matrix evolves over the study period. In the beginning it is highly negatively skewed, with a predominance of distances $d_{ij}$ of 0.9 to 1. Over time it becomes nearly symmetrical, with a mode of about 0.6 for the distances in the post 2000}.
% to confort this nice economy and finance interpretation and comments may be it is needed to say more more about the change structure of the graphs  before and after but hard without final figs.
\\
\indent
The change point located in 2007 is also characterized by a decrease in distance and an increase in connectivity in the graph topologies. 
Eccentricity however decreases here as much as the other distance indicators (Fig. \ref{fig:figb} (b)). 
A possible economic interpretation for the date of this change may be related to the beginning of the financial crisis. 
Another less trivial interpretation is that the crisis also corresponds to a return of North-South capital flows. 
% please say where we see this on graphs. Pls related comments as much as can be with figures and observations on graphs. 
This return precisely results from low capital returns in advanced economies in the context of a long-lasting economic slowdown. 
This economic interpretation is notably consistent with the large decrease of the eccentricity {after the 2007 change point} (as opposed to the 1983 change point), highlighting that the joint convergence of developing and advanced economies accelerates in the post 2007 period. % {\bf as well as the convergence of advanced economies}.  
MST representations also show that the USA do not maintain their 1984-89 central position (Fig. \ref{fig:figc} (c)), and that the global structure seen in Fig. \ref{fig:figc} (d) is a core of highly integrated nodes, with a ring  
of emerging economies that are mainly located (with MDS) in the same peripheral area -- whereas in the previous representation they are located in the peripheral areas of several advanced economies. 
%{\bf It is also interesting to notice in the last subperiod MDS plane representations which take into account the whole distance matrices and not only the sparse MST matrices that emerging economies are mainly located in the same peripheral area whereas in the former representations they were located in the peripheral areas of several advanced economies (Fig. \ref{fig:figc} upper right)}. 

\indent The change point located in 1964 is, we believe, mostly related to the structure of the data. 
Data are missing for a large number of countries before 1964, and it is thus likely that this change in the graph structures is mostly driven by the incorporation in the analysis of several new countries.

\subsection{Refined segmentation.} Decreasing regularization parameter $\lambda$  to {50}  produces a segmentation in a larger number of eras, as expected.
Fig. \ref{fig:figb} right indicates that decreasing $\lambda$ results into a further splitting of the 4-era obtained with a larger $\lambda$, hence yielding a \emph{hierarchical segmentation}, a satisfactory outcome, not a priori imposed or induced by the analysis procedure, hence really associated to the very structure of the data. 
The 1964-1983 period is split into 4 different subperiods. 
This can be related to the large number of major economic and financial events occurring during the 1970s, such as the end of the dollar/gold convertibility in 1971, the first and second oil shocks in 1973 and 1979, the end of the Bretton Woods system in 1976 and the radical shift in the monetary policy stance of the United States in 1979. 
For 1983-2007, it is interesting to observe that the sub-segmentation highlights a local increase in the eccentricity measure {in 1991} while the overall trend of distance indicators is decreasing {over the whole study period}. 
This is consistent with this peculiarity of the subperiod, when developing economies were left outside of international capital flows.

\section{Conclusions and perspectives}
\label{sec:conc}

The present contribution has shown a segmentation methodology, combining covariance graph constructions and analysis and state-of-the-art optimization based segmentation along time. 
It allows to delineate international financial integration eras using network indicators which is, to the best of our knowledge, an original contribution to the literature. 
%This methodology leads to topological representations and stable integration eras that enrich our economic interpretations. 

An important finding is that the analysis suggests that the main cause of the segmentation is the pattern of international capital flows. 
The existence or not of truly globalized capital flows, i.e., including large North-South flows, seems crucial. 
Regulatory shocks, in particular the end of the Bretton Woods system in 1976, appear as secondary changes in subperiod quasi-hierarchical segmentation, achieved by decreasing the penalization parameter $\lambda$.
This result is particularly interesting insofar as the segmentation method does not necessarily induce a re-segmentation of the main segmentation when the penalization parameter decreases. Moreover, the re-segmentation precisely occurs in the 70s, where the two possible periodizations of globalization eras are superimposed.

Finally this original result is also interesting as the regulatory environment of international monetary relationships is usually considered as the most relevant delimitation of globalization eras, while the key role of the patterns of international capital flows is under-estimated. This suggests initial responses to a set of outstanding issues. Among these outstanding issues is whether the deep changes in monetary and financial institutional regimes would be the product of the changes in the structure of international capital movements. In other words, have the major institutional changes in the world since 1960 around the key break dates (1971, 1973, 1976) been dictated by changes in the structure of global financial networks? Is international financial integration the driving force of institutional change? 

In future work, we plan to test the robustness of our conclusions by using various correlations measures, network representations, and indices of the graph topology.

\section{Acknowledgment}
This work was conducted in the CAC (Cliometrics \& Complexity group) at IXXI (Institut Rh\^onalpin des Syst\`emes Complexes), and supported by the GRAPHSIP project ANR-14-CE27-0001-02, and by the ACADEMICS Grant given by the IDEXLYON project of the Universit\'{e} de Lyon, as part of the ``Programme Investissements d'Avenir" ANR-16-IDEX-0005.

% Biblio : initiale prenom ? 

%\clearpage
\newpage

\bibliographystyle{elsarticle-num.bst}
%\bibliography{biblio.bib}
% \bibliographystyle{unsrtnat}
\bibliography{Bib_endog_eras} 

\end{document}